\documentclass[aip,reprint]{revtex4-1}
\usepackage[T1,T2A]{fontenc}
\usepackage[utf8]{inputenc}
\usepackage[russian,english]{babel}
\usepackage{graphicx}
\usepackage{dcolumn}
\usepackage{bm}
\usepackage{graphicx}
\usepackage{amsmath}
\usepackage{amssymb}
\usepackage{pifont}
\usepackage{epstopdf}
\usepackage{color}

\begin{document}

\title{Limiting propulsion of ionic microswimmers}
\author{Evgeny S. Asmolov (Евгений С. Асмолов)}
\affiliation{Frumkin Institute of Physical Chemistry and
Electrochemistry, Russian Academy of Science, 31 Leninsky Prospect,
119071 Moscow, Russia}
\affiliation{Institute of Mechanics, Lomonosov Moscow State
University, 1 Michurinskiy Prospect, 119991 Moscow, Russia}
\email[Corresponding author: ]{aes50@yandex.ru}

\author{Olga I. Vinogradova (Ольга И. Виноградова)}
\affiliation{Frumkin Institute of Physical Chemistry and
   Electrochemistry, Russian Academy of Science, 31 Leninsky Prospect,
   119071 Moscow, Russia}

\begin{abstract}
Self-propulsion of catalytic Janus swimmers in electrolyte solutions is
induced by inhomogeneous ion release from their surface. Here, we consider
the experimentally relevant cases of particles which emit only one type of
ions (type I) or equal fluxes of cations and anions (type II). In the limit
of a thin electrostatic diffuse layer we derive a nonlinear outer solution
for the electric field and concentrations of active (i.e. released from the
surface) and passive ionic species. We show that for swimmers of type I both the maximum ion flux and propulsion velocity are constrained. This suggests that the propulsion of Janus swimmers can be optimized by tuning the concentration of active
ions.
\end{abstract}

\maketitle

\section{Introduction}

Catalytic Janus swimmers have drawn a considerable interest during last decades%
\cite{moran2017,dey2017,asmolov2022COCIS} because of their ability to move
autonomously. The potential applications include drug delivery,
lab-on-a-chip devices and nano-robotics\cite{hu2020micro}. The
self-propulsion mechanism of ionic swimmers is based on an inhomogeneous
ion release at the particle surface.~Ion diffusion induces concentration
gradient and electric field\cite{golestanian2005propulsion}.

\begin{figure}[h]
\centering
\includegraphics[width=1\columnwidth]{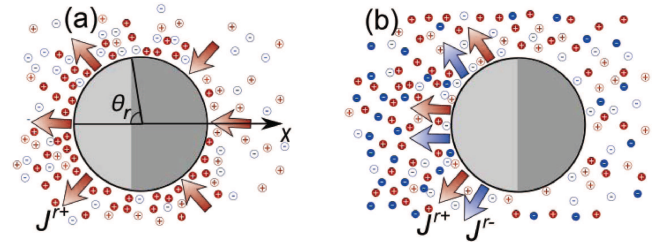}
\caption{Two types of ionic swimmers. (a) Type I releases cations only at
the active side and adsorb them at the other side. (b) Type II releases both
cations and anions at the active side. }
\label{sketch}
\end{figure}

Despite the variety of catalytic reactions, two main models for ionic flux
are accepted\cite{asmolov2022COCIS,Wang2020practical,peng2022generic} (see
Fig.~\ref{sketch}): only one type of ions is released (Type I) and equal
fluxes of cations and anions (Type II). The classical examples of the first
type include bimetallic\cite{paxton2004} and metallic-insulator \cite%
{howse2007} Janus particles. As a result of hydrogen peroxide decomposition
on the catalytic particle side, protons are released into solution and then
are adsorbed on the non-catalytic side. The ionic flux for the second type
can be generated by the salt dissolution in water\cite{mcdermott2012},
enzyme-enhanced decomposition of organic compounds\cite%
{dey2015micromotors,patino2018fundamental} or by photochemical reactions\cite%
{zhou2018}.

Charged particle immersed in an electrolyte solution is surrounded by an
electric diffuse layer (EDL), where charge imbalance is large and the
electric field is strong. Concentration of ionic species, electric potential
and fluid flow near catalytic swimmers are governed by a system of the
Nernst-Planck, Poisson and Stokes equations. The EDL is thin for swimmers of
micron size, and the method of matched asymptotic expansions is usually
applied for the theoretical description of self-propulsion\cite%
{prieve.dc:1984,anderson.jl:1989,golestanian2007}. The EDL effect appears
macroscopically as a slip velocity at the particle surface. The electric
field in the outer region (at distances of the order of the particle size)
is not zero due to ion diffusion and small charge imbalance and can effect
on the slip velocity\cite{nourhani2015,asmolov2022self}. The analytical
solutions in the outer region has been obtained for the swimmer of Type I%
\cite{nourhani2015,ibrahim2017} under the assumption that field disturbances
are small and the Nernst-Planck equations can be linearized.

Recently, we have proposed\cite{asmolov2022self,asmolov2022MDPI} non-linear
analytical solutions to Nernst-Planck and Poisson equations for the cases
when bulk electrolyte contains active (released from the swimmer surface)
ions only. However, the self-propulsion in experiments is carried out in
media containing both active and passive ions. The addition of passive salt
reduces the swimmer velocity\cite{moran2017,ibrahim2017,arque2020ionic}. In
the present paper, we extend our previous solutions to include the effect of
passive ions in the bulk. We obtain the analytical solutions in the outer region
for both swimmer types and all ionic species. The depletion of active ions
arise for the swimmer of Type I near an adsorbing particle side. This
imposes a constraint on the maximum swimmer velocity.

Our paper is organized as follows. The governing equations are formulated in
Sec.~\ref{s2}. Their solution at distances of the order of particle size
(outer region) is given in Sec.~\ref{s3}. Our numerical results for
concentrations and for the limiting particle velocity are presented in Sec.~%
\ref{res} and are summarized in Sec.~\ref{s5}.

\section{Governing equations\label{s2}}

We consider an 1:1 aqueous electrolyte solution with the bulk ion
concentration $c_{\infty }.$ It includes passive salt ions with
dimensionless concentrations $C^{s\pm }\left( \mathbf{r}\right) \ $(scaled
by $c_{\infty })$ and ions which are released from the particle surface with
concentrations $C^{r\pm }\left( \mathbf{r}\right) $. Here the upper (lower)
sign corresponds to the positive (negative) ions. The Peclet number is
assumed to be small, so that the convective fluxes are ignored, and the ion
fluxes are governed by dimensionless Nernst-Planck equations
\begin{equation}
\nabla \cdot \mathbf{J}^{i}=0,  \label{NPH}
\end{equation}%
\begin{equation}
\mathbf{J}^{i}=-\nabla C^{i}\mp C^{i}\nabla \Phi .  \label{NP1}
\end{equation}%
Here $i=r\pm ,$ $s\pm ,$ the electric potential $\Phi $ is scaled by $%
k_{B}T/e,$ where $e$ is the elementary positive charge, $k_{B}$ is the
Boltzmann constant, $T$ is the temperature of the system, and coordinates
are scaled by the particle size $a.$

Unequal distributions of oppositely charged ions generate an electric field,
governed by the Poisson equation for the electric potential,%
\begin{equation}
\Delta \Phi =\lambda ^{-2}\frac{C^{s+}+C^{r+}-C^{s-}-C^{r-}}{2},
\label{pois1}
\end{equation}%
where the dimensionless parameter $\lambda =\left( \kappa a\right) ^{-1}$ is
the ratio of the Debye length $\kappa ^{-1}=\left( 8\pi e^{2}c_{\infty
}/\epsilon k_{B}T\right) ^{-1/2}$ to the particle size, $\epsilon $ is the
permittivity of the fluid. We consider large particles with a thin electric
double layer (EDL), i.e. $\lambda \ll 1,$ and construct the asymptotic
solution using the method of matched expansions in two regions with
different lengthscales. The lengthscale of the outer region is the particle
size $a$ and that for the inner region is the Debye length $\kappa ^{-1}$.

The boundary conditions at the particle surface $S$ (defined by $\mathbf{r}=%
\mathbf{r}_{s}$) read,
\begin{gather}
\mathbf{r}\in S:\ \mathbf{J}^{r+}\cdot \mathbf{n}=-\mathrm{Da}j^{+}\left(
\mathbf{r}_{s}\right) ,  \label{sp1} \\
\mathbf{J}^{r-}\cdot \mathbf{n}=-\mathrm{Da}j^{-}\left( \mathbf{r}%
_{s}\right) ,  \label{sp2} \\
\mathbf{J}^{s\pm }\cdot \mathbf{n}=0,  \label{sps} \\
\ \Phi =\phi _{s}\left( \mathbf{r}_{s}\right) ,  \label{sp3} \\
\mathrm{Da}=\frac{Ja}{c_{\infty }D^{r+}},  \label{Da_def}
\end{gather}%
where $\mathrm{Da}$ is the Damk\"{o}hler number and $J$ is the average module of
dimensional flux of released cations. The first two conditions define the
surface fluxes of active ions, and the third one the zero fluxes for passive
ions. The functions $j^{\pm }\left( \mathbf{r}_{s}\right) $ characterize
dimensionless inhomogeneous ion productions at the particle surface. Because
of definition (\ref{Da_def}) the cation flux satisfies the condition
\begin{equation}
\frac{1}{S}\int_{S_{p}}\left\vert j^{+}\right\vert dS=1.  \label{norm0}
\end{equation}%
The net flux for the swimmers of Type I should vanish in a steady state:
\begin{equation}
\int_{S_{p}}j^{+}dS=0.  \label{norm1}
\end{equation}%
The boundary condition (\ref{sp3}) sets the surface potential $\phi
_{s} $ which are assumed to be constant.

Boundary conditions at infinity are,%
\begin{equation}
r\rightarrow \infty :\ C^{s\pm }=C_{\infty }^{s\pm },\ C^{r\pm }=C_{\infty
}^{r\pm },\ \Phi =0.  \label{bc_f_inf}
\end{equation}%
The bulk electrolyte solution is electroneutral, so that $$C_{\infty
}^{s+}+C_{\infty }^{r+}=C_{\infty }^{s-}+C_{\infty }^{r-}=1.$$

The fluid flow satisfies the Stokes equations,
\begin{equation}
\mathbf{\nabla \cdot v}=0\mathbf{,\quad }\Delta \mathbf{v}-\mathbf{\nabla }p=%
\mathbf{f}.  \label{NS}
\end{equation}%
Here $\mathbf{v}$ and $p$ are the dimensionless fluid velocity and pressure
(scaled by $\dfrac{\epsilon k_{B}^{2}T^{2}}{4\pi \eta e^{2}a}$ and $\dfrac{%
\epsilon k_{B}^{2}T^{2}}{4\pi e^{2}a^{2}}$), $\eta $ the viscosity and $%
\mathbf{f}=-\Delta \Phi \mathbf{\nabla }\Phi $ is the electrostatic body
force.

\section{Theory}

\label{s3}

In this section we present the solution of the system (\ref{NPH})-(\ref%
{Da_def}) in the outer region with the lengthscale $a.$ Since $\lambda \ll 1,
$ the leading-order solution of (\ref{pois1}) in the outer region is
\begin{equation}
C^{s+}+C^{r+}=C^{s-}+C^{r-}=C\left( \mathbf{r}\right) ,  \label{cnet}
\end{equation}%
i.e. the electroneutrality holds to $O\left( \lambda ^{2}\right) .$ However,
one should take into account the small charge $C^{s+}+C^{r+}-C^{s-}-C^{r-}=O%
\left( \lambda ^{2}\right) $, since it induces a finite potential difference
in the outer region. We sum up the equations (\ref{NPH}) for
positive and negative ions, and the system then becomes
\begin{eqnarray}
\Delta C+\nabla \cdot \left( C\nabla \Phi \right)  &=&0,  \label{NP2} \\
\Delta C-\nabla \cdot \left( C\nabla \Phi \right)  &=&0.  \label{NP3}
\end{eqnarray}%
Summing up and subtracting equations (\ref{NP2}) and (\ref{NP3}) we obtain%
\cite{nourhani2015,ibrahim2017,asmolov2022MDPI}:
\begin{eqnarray}
\Delta C &=&0,  \label{dif} \\
\nabla \cdot \left( C\nabla \Phi \right)  &=&0.  \label{pot2}
\end{eqnarray}%
Likewise, by calculating the sum and difference of Eqs.(\ref{sp1}), (\ref%
{sp2}) and (\ref{sps}) we derive the boundary conditions to (\ref{dif}) and (%
\ref{pot2}),
\begin{gather}
\mathbf{r}\in S:\ \mathbf{\nabla }C\cdot \mathbf{n}=-\frac{\mathrm{Da}}{2}%
\left( j^{+}+j^{-}\right) ,  \label{il1} \\
C\mathbf{\nabla }\Phi \cdot \mathbf{n}=-\frac{\mathrm{Da}}{2}\left(
j^{+}-j^{-}\right) .  \label{il2}
\end{gather}%
Solution of Laplace equation (\ref{dif}) makes it possible to find the net
ion concentrations $C$. However, the solution of a non-linear equation for
the electric potential \eqref{pot2} is generally not straightforward. We
consider here the case of the constant ratio of the surface fluxes,
\begin{equation}
j^{-}\left( \mathbf{r}_{s}\right) /j^{+}\left( \mathbf{r}_{s}\right) =\alpha
=\text{const}.  \label{fl_s}
\end{equation}%
The condition is valid for the two swimmer type: $\alpha =0$ for the Type I
(cations are released only) and $\alpha =D^{r+}/D^{r-}$ for the Type II. The
solution of (\ref{pot2}) satisfying the boundary conditions (\ref{bc_f_inf})
and (\ref{il2}) is\cite{asmolov2022MDPI}%
\begin{equation}
\Phi =-\beta \ln \left( C\right) ,  \label{phi_o2}
\end{equation}%
\begin{equation}
\beta =\dfrac{\alpha -1}{\alpha +1}.  \label{beta}
\end{equation}%
The above dimensionless parameter is limited, namely $\left\vert \beta
\right\vert \leq 1.$

To find the concentrations $C^{i}$ we do not need to solve the system (\ref%
{NPH})-(\ref{sps}). The boundary fluxes for passive ions $C^{s\pm }$ are
zero, so their fields obey the Boltzmann distribution,%
\begin{equation}
C^{s\pm }=C_{\infty }^{s\pm }\exp \left( \mp \Phi \right) =C_{\infty }^{s\pm
}C^{\pm \beta }.  \label{csp}
\end{equation}%
Once $C$ and $C^{s\pm }$ are known we can also determine the distributions
of active ions from Eqs. (\ref{cnet}), (\ref{csp}):%
\begin{eqnarray}
C^{r+} &=&C-C^{s+}=C-C_{\infty }^{s+}C^{\beta },  \label{crp} \\
C^{r-} &=&C-C^{s-}=C-C_{\infty }^{s-}C^{-\beta }.  \label{crm}
\end{eqnarray}

Thus, we obtain concentrations of all ionic species in the outer region for
arbitrary $\mathrm{Da}$ and $j^{\pm }\left( \mathbf{r}_{s}\right) .$ This is
one of the main results of our work. However, physically reasonable results
may not be found for any boundary conditions. The restriction comes from the
condition that the concentrations may not be negative for all $\mathbf{r}$
and $i,$%
\begin{equation}
C^{i}\left( \mathbf{r}\right) \geq 0.  \label{constr}
\end{equation}%
This implies physical constraints on $\mathrm{Da}$ for given $C_{\infty }^{i}
$ and $j^{\pm }\left( \mathbf{r}_{s}\right) $. When the normal gradient $%
\mathbf{\nabla }C\cdot \mathbf{n}$ is negative over the entire particle
surface, i.e. the net flux of released ions $j^{+}\left( \mathbf{r}%
_{s}\right) +j^{-}\left( \mathbf{r}_{s}\right) $ is always positive as for
the Type II, the solution of (\ref{dif}) satisfies the condition $C\geq 1$
for any $\mathbf{r,}$ i.e. the net concentration exceeds the bulk value.
This means that the concentration of released ions (Eqs. (\ref{crp}), (\ref%
{crm})) is always positive since $C_{\infty }^{s\pm }\leq 1.$ The opposite
situation takes place for the Type I when ions are adsorbed at some surface
portion, where the surface flux is negative and $\mathbf{\nabla }C\cdot
\mathbf{n}$ is positive. Then we may obtain physically impossible solution $%
C\leq 0,$ when $\mathrm{Da}$ is large enough. Moreover, negative solutions
for active ions can follow from Eqs. (\ref{crp}), (\ref{crm}) even at
smaller $\mathrm{Da}$ (see Sec. \ref{res}). Therefore, there is a limiting
value $\mathrm{Da}_{\mathrm{max}}$ for which the minimum concentration $%
C_{\infty }^{r\pm }=0.$

Solution in the inner region expresses the slip velocity $\mathbf{v}_{s}$ at
the outer edge of the EDL\cite{prieve.dc:1984,anderson.jl:1989}
\begin{equation}
\mathbf{v}_{s}=\psi \nabla _{s}\Phi _{s}+4\nabla _{s}(\ln C_{s})\ln \left[
\cosh \left( \frac{\psi }{4}\right) \right] ,  \label{slip}
\end{equation}%
\begin{equation*}
\psi =\phi _{s}-\Phi _{s},\quad \Phi _{s}=\Phi \left( \mathbf{r}_{s}\right)
,\quad C_{s}=C\left( \mathbf{r}_{s}\right) .
\end{equation*}%
Here $\nabla _{s}$ is the gradient operator along the particle surface, the
first and the second terms in Eq.\eqref{slip} are associated with electro-
and diffusio-osmotic flows, respectively, $\psi $ is the potential drop in
the inner region, $\Phi _{s}$ and $C_{s}$ are the outer potential and the
concentration at the dividing surface between the inner and outer regions.

The velocity of a freely moving in the $x-$direction particle can be
determined by using the reciprocal theorem\cite{teubner1982}
\begin{equation}
v_{p}=-\frac{1}{6\pi }\int_{V_{f}}\mathbf{f}\cdot \left( \mathbf{v}_{1}-%
\mathbf{e}_{x}\right) dV.
\end{equation}%
The integral is evaluated over the whole fluid volume $V_{f}$ and $\mathbf{v}%
_{1}\left( \mathbf{r}\right) $ represents the velocity field for the
particle of the same shape that translates with the velocity $\mathbf{e}_{x}$
in a stagnant fluid. The velocity can be presented as a superimposition of
the contributions of the outer and inner regions\cite{asmolov2022self},
\begin{equation}
v_{p}=v_{po}+v_{pi},  \label{v_full}
\end{equation}%
\begin{equation}
v_{po}=-\frac{\beta ^{2}}{6\pi }\int_{V_{f}}\frac{\left( \mathbf{\nabla }%
C\right) ^{2}\mathbf{\nabla }C\cdot \left( \mathbf{v}_{1}-\mathbf{e}%
_{x}\right) }{C^{3}}dV,  \label{v_o}
\end{equation}%
\begin{equation}
v_{pi}=-\frac{1}{S}\int_{S_{p}}\left( \mathbf{v}_{s}\cdot \mathbf{e}%
_{x}\right) dS.  \label{v_pedl}
\end{equation}%
Equations (\ref{slip})-(\ref{v_pedl}) do not include the concentrations of
each ionic species, but only the net value $C.$ Hence, the velocity is
independent of $C_{\infty }^{s+}$ provided $\mathrm{Da}<\mathrm{Da}_{\mathrm{%
max}}\left( C_{\infty }^{s+}\right) .$

When the Damk\"{o}hler number is small, the concentration and potential fields
are slightly disturbed in the outer region, $C=1+O\left( \mathrm{Da}\right)
\mathrm{,\ }\Phi =O\left( \mathrm{Da}\right) ,$ $\psi \simeq \phi _{s}$ as $%
\mathrm{Da}\ll 1.$ Contribution of the outer region to the velocity is $%
O\left( \mathrm{Da}^{3}\right) $ and can be neglected, since the integrand
in Eq. (\ref{v_o}) is proportional to $\left( \mathbf{\nabla }C\right) ^{3}$%
. Then the dominant contribution of the order of $\mathrm{Da}$ comes from
the inner region (\ref{v_pedl}) with the linearized slip velocity:%
\begin{equation}
\mathbf{v}_{s}=\phi _{s}\nabla _{s}\Phi _{s}+4\nabla _{s}C_{s}\ln \left[
\cosh \left( \frac{\phi _{s}}{4}\right) \right] =O\left( \mathrm{Da}\right) .
\label{vs_lin}
\end{equation}%
Below we show that this linear solution approximates well the exact results
up to $\mathrm{Da}\sim 1.$

\section{Results and discussion}

\label{res}

We consider spherical particle with symmetric ion release with respect to $%
x- $ axis, so that $j^{\pm }$ depends on azimuthal angle $\theta $ only. The
boundary condition (\ref{il1}) is rewritten as%
\begin{equation}
r=1:\ \partial _{r}C=-\frac{\mathrm{Da}}{2}\left( j^{+}+j^{-}\right) .
\label{bc_p}
\end{equation}

We use the spherical coordinate system to solve Eq. (\ref{dif}) with the
boundary conditions (\ref{bc_p}) and present the solution in terms of
Legendre polynomials $P_{n}\left( \cos \theta \right) $,
\begin{equation}
C=1+\frac{\mathrm{Da}}{2}\xi \left( r,\theta \right) ,  \label{c_sol}
\end{equation}%
\begin{eqnarray}
\xi &=&\sum_{n=0}^{\infty }\frac{j_{n}}{n+1}P_{n}\left( \cos \theta \right)
r^{-n-1},  \label{qn} \\
j_{n} &=&\left( n+\frac{1}{2}\right) \int_{0}^{\pi }\left( j^{+}
+j^{-}\right) P_{n} \sin \theta d\theta .  \label{qn1}
\end{eqnarray}

We model inhomogeneous distributions $j^{+}\left( \theta \right)
,j^{-}\left( \theta \right) $ by piecewise constant functions, i.e. the ions
are released from the surface for $\theta \leq \theta _{r}.$ At the other
sphere side, $\theta _{r}<\theta \leq \pi ,$ cations are adsorbed for the
swimmer Type I (see Fig. \ref{sketch}(a)), and the fluxes are zero for the
Type II (see Fig. \ref{sketch}(b)). Then the surface flux distributions
satisfying conditions (\ref{norm0}) and (\ref{norm1}) are
\begin{equation}
\begin{array}{cc}
0\leq \theta \leq \theta _{r}: & j^{+}=\frac{1}{2\left( 1+\cos \theta
_{r}\right) }<0,\ j^{-}=0, \\
\theta _{r}<\theta \leq \pi : & j^{+}=-\frac{1}{2\left( 1-\cos \theta
_{r}\right) }>0,\ j^{-}=0,%
\end{array}
\label{j1}
\end{equation}%
for the first type, and
\begin{equation}
\begin{array}{cc}
0\leq \theta \leq \theta _{r}: & j^{+}=j^{-}/\alpha =\frac{1}{\left( 1+\cos
\theta _{r}\right) }>0, \\
\theta _{r}<\theta \leq \pi : & j^{+}=j^{-}=0,%
\end{array}
\label{j2}
\end{equation}%
for the second one.

\begin{figure}[h]
\centering
\includegraphics[width=0.9\columnwidth]{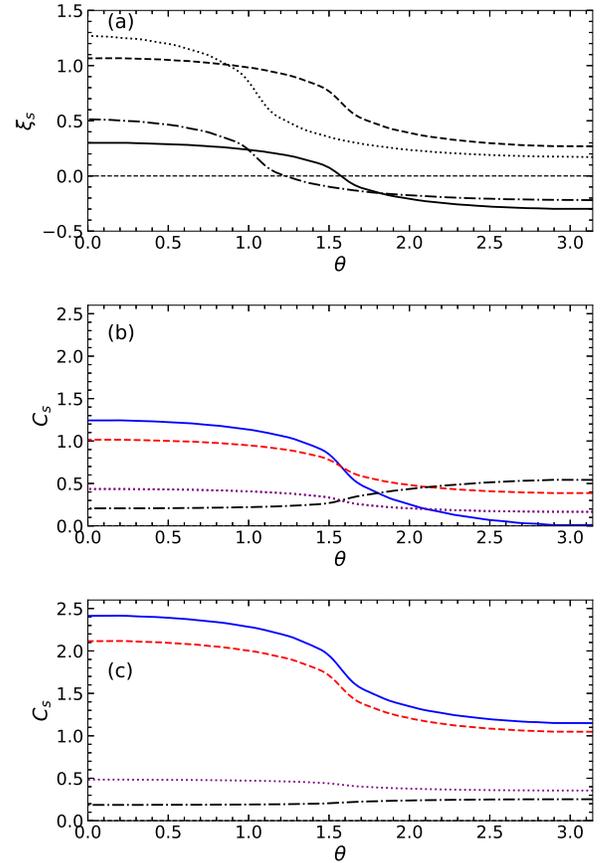}
\caption{(a) Reduced ion concentration at the particle surface as a function
of $\protect\theta $. Solid and dashed-dotted lines correspond to the
swimmer Type I with $\protect\theta _{r}=\protect\pi /2$ and $\protect\theta %
_{r}=\protect\pi /3$, respectively, dashed and dotted lines to the Type II, $%
\protect\theta _{r}=\protect\pi /2$ and $\protect\theta _{r}=\protect\pi /3$%
. Concentrations of ion species for swimmers of (b) the Type I and (c) Type II
calculated using $\protect\theta _{r}=\protect\pi /2$, $\mathrm{Da}=3$ and $%
C_{\infty }^{s+}=0.3$. Solid and dashed lines are concentrations of released
cations and anions, respectively, dashed-dotted and dotted lines are
concentrations of salt cations and anions. }
\label{cs}
\end{figure}

We evaluate numerically the integrals in Eq. (\ref{qn1}) to obtain a reduced
ion concentration $\xi \left( r,\theta \right) $ with the surface fluxes (%
\ref{j1}) and (\ref{j2}) on a uniform grid in $\theta $\ with $N_{\theta
}=200.$ To evaluate the particle velocity we calculate the surface integral
in Eq. \eqref{v_pedl}, and the volume integral in Eq. \eqref{v_o} using the
same grid in $\theta $\ and a non-uniform grid in $r$\ (with a grid step
varying as $r^{2}$) with $N_{r}=100$\ nodes and a cut-off radius $R_{out}=100
$. Figure \ref{cs}$\left( a\right) $ shows the functions $\xi _{s}\left(
\theta \right) =\xi \left( 1,\theta \right) $ for the swimmer of the Type I
with $\beta =-1$ (solid and dashed-dotted lines) and Type II with $\beta =-0.5,\ \alpha
=D^{r+}/D^{r-}=1/3$ (dashed and dotted lines). The functions decay with $\theta $, i.e.
the ion concentration is always greater at the catalytic section where ions
are released. For the Type II, $\xi _{s}>0$ for any angle $\theta _{r}$ over
the entire surface, so that $C\geq 1$ (see Eq. (%
\ref{c_sol})), i.e. the net concentration is always greater than the bulk
value. In contrast, for the Type I, $\xi
_{s}\left( \theta \right) $ changes the sign so that $C_{s}<1$ at the rear
sphere side where the ions are adsorbed. We obtain from Eq. (\ref{c_sol})
non-physical values $C_{s}<0$ when $\mathrm{Da>Da}_{\mathrm{max}}=-2/\xi _{%
\mathrm{min}}$ where $\xi _{\mathrm{min}}=\xi \left( 1,0\right) <0$ is the
absolute minimum of the function $\xi $ attained at the rear sphere pole.
The minimum $\xi _{\mathrm{min}}$\textrm{\ }grows and the limiting Damk\"{o}hler
number decreases with $\theta _{r}$ (smaller area of cation adsorption).

The limiting value of $\mathrm{Da}$ becomes even smaller when we take into
account the condition that the concentrations of all ionic species should
also be positive. We evaluate $C^{i}$ by using Eqs. (\ref{csp})-(\ref{crm})
and present the results in Fig. \ref{cs}$\left( b,c\right) .$ For the Type
II, Eqs. (\ref{csp})-(\ref{crm}) predict positive concentrations for all
species (see Fig. \ref{cs}$\left( c\right) )$ as $C\geq 1,$ $C_{\infty
}^{s+}\leq 1,\ \left\vert \beta \right\vert \leq 1,$ so that there is no
limitation on $\mathrm{Da.}$ The concentrations of passive ions
(dashed-dotted and dotted lines) are rather small, while those for the
released ions (solid and dashed lines) are large and surpass the bulk value $%
C_{\infty }^{r+}=0.7.$ Another behavior can be viewed in Fig. \ref{cs}$%
\left( b\right) $ for the Type I. The concentration of released cations
(solid line) is close to zero at $\theta =\pi .$ Therefore, the Figure
illustrates the limiting regime which is attained at $\mathrm{Da}_{\mathrm{%
max}}\simeq 6$ for given $C_{\infty }^{r+}$ and $j^{+}\left( \theta \right)
. $ Greater $\mathrm{Da}$ is impossible since Eq. (\ref{crm}) gives $%
C^{r+}<0.$ We note that $C_{s}^{r+}\left( \theta \right) $ (solid line)
decreases significantly along the sphere surface while the concentration of
passive cations $C_{s}^{s+}\left( \theta \right) $ (dash-dotted line) grows,
so that the variation of the net concentration remains not too large. Thus
the presence of passive salt reduces the limiting $\mathrm{Da.}$

\begin{figure}[h]
\centering
\includegraphics[width=0.87\columnwidth]{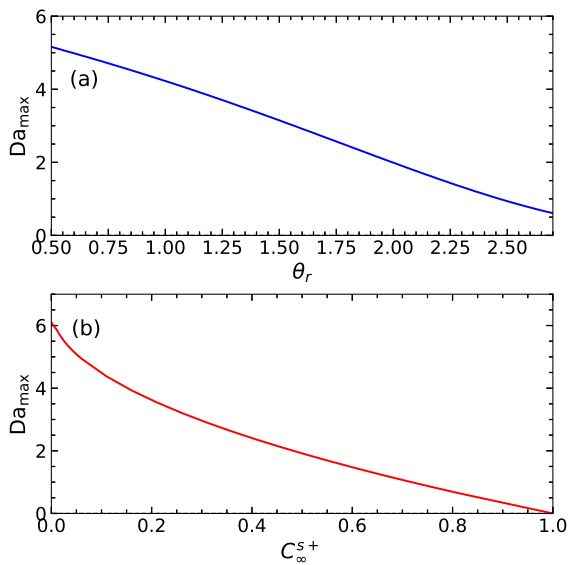}
\caption{Limiting Damk\"{o}hler number calculated as functions (a) of $\protect%
\theta _{r}$ for $C_{\infty }^{s+}=0.3$ and (b) of $C_{\infty }^{s+}$ for $%
\protect\theta _{r}=\protect\pi /2$. }
\label{damax}
\end{figure}

Condition $C^{r+}\geq 0$ can be reformulated as the limitations of the
minimum bulk value or of the maximum Damk\"{o}hler number. Equation (\ref{crp})
with $\beta =-1$ reads
\begin{equation}
C^{r+}=C-C_{\infty }^{s+}/C\geq 0.  \label{ch1}
\end{equation}%
Then by using (\ref{c_sol}) we obtain
\begin{equation}
C_{\infty }^{s+}\leq C_{\min }^{2}=\left( 1+\frac{\mathrm{Da}}{2}\xi _{%
\mathrm{min}}\right) ^{2},\ C_{\infty }^{r+}\geq 1-C_{\min }^{2}> 0,
\label{con1}
\end{equation}%
or equivalently,
\begin{equation}
\mathrm{Da}<\mathrm{Da}_{\mathrm{max}}=2\frac{\sqrt{C_{\infty }^{s+}}-1}{\xi
_{\mathrm{min}}}.  \label{dam}
\end{equation}%
The maximum Damk\"{o}hler number depend only on the salt concentration $%
C_{\infty }^{s+}$ and the angle $\theta _{r}$. Figure \ref{damax}$%
\left( a,b\right) $ shows $\mathrm{Da}_{\mathrm{max}}$ as the functions of
the two parameters. It decreases with $\theta _{r}$ and with $C_{\infty
}^{s+}.$ The active cations cannot released on the catalytic surface ($%
\mathrm{Da}_{\mathrm{max}}\rightarrow 0$ and $v_{p}\rightarrow 0)$ when they
are absent in the bulk, $C_{\infty }^{r+}\rightarrow 0,$ or when the area of
cation adsorption tends to zero $\left( \theta _{r}\rightarrow \pi \right) .$%

\begin{figure}[h]
\centering
\includegraphics[width=0.95\columnwidth]{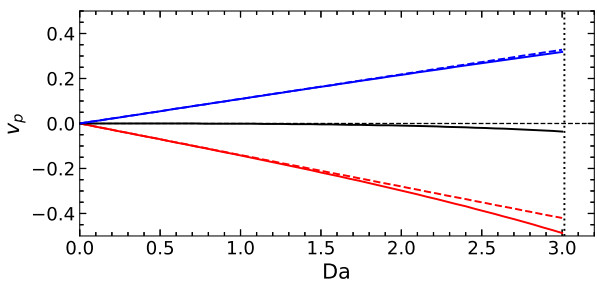}
\caption{Particle velocities calculated using $\protect\phi _{s}=-1$, $0$, $%
1 $ (from top to bottom), $\protect\theta _{r}=\protect\pi /2$, $C_{\infty
}^{s+}=0.3$ as functions of $\mathrm{Da}$. Solid curves are the total
velocities $v_{pi}+v_{po}$, dashed lines correspond to the linear solution (%
\protect\ref{vs_lin}), dotted line is the limiting Damk\"{o}hler number. }
\label{vpda}
\end{figure}

The particle velocity $v_{p}$ depends on three dimensionless parameter, $%
\mathrm{Da,}$ $\phi _{s},$ $\theta _{r}.$ Equations (\ref{slip})-(\ref%
{v_full}) do not include the concentrations of each ionic species, but only
the net value $C,$ so the velocity remains the same for any $C_{\infty }^{s+}
$ when $\mathrm{Da}<\mathrm{Da}_{\mathrm{max}}\left( C_{\infty }^{s+},\theta
_{r}\right) .$ We evaluated $v_{p}$ earlier for the swimmer Type II\cite%
{asmolov2022self} (see also Fig. 4 in \cite{asmolov2022COCIS}). $\mathrm{Da}$
can be arbitrary in this case, and $v_{p}$ grows logarithmically as $\mathrm{%
Da}\rightarrow \infty .$

We show the particle velocity for the swimmer Type I in Fig. \ref{vpda} as
function of the Damk\"{o}hler number for different $\phi _{s}.$ The velocity of
an uncharged particle $\left( \phi _{s}=0\right) $ is very small, and at
finite $\phi _{s}$ and $\mathrm{Da,}$ it is close to the predictions of the
linear solution (\ref{vs_lin})$\mathrm{.}$ The maximum velocity, that is
reached at $\mathrm{Da}=\mathrm{Da}_{\mathrm{max}}\left( C_{\infty
}^{s+},\theta _{r}\right) ,$ strongly depends on $C_{\infty }^{s+}$.

\begin{figure}[h]
\centering
\includegraphics[width=1\columnwidth]{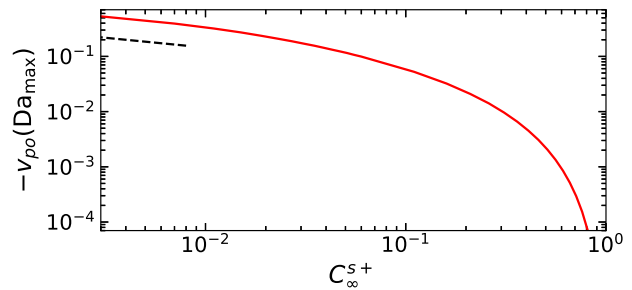}
\caption{The contribution of the outer region to the particle velocity $%
v_{po}$ calculated for $\mathrm{Da}=\mathrm{Da}_{\mathrm{max}}$ and $\protect%
\theta _{r}=\protect\pi /2$ (solid line). Dashed line is the power function
with the exponent $-0.355$. }
\label{vomax}
\end{figure}

We further study the dependences of the maximum particle velocity $%
v_{p}\left( \mathrm{Da}_{\mathrm{max}}\right) $ on $C_{\infty }^{s+},$ $%
\theta _{r}$ and $\phi _{s},$ and evaluate separately the contributions of
the outer and the inner regions, $v_{po}$ and $v_{pi}$. The velocity $v_{po}$
is shown in Fig. ~\ref{vomax} as the function of the salt concentration. It
is negative for any $C_{\infty }^{s+},$ and is extremely small at finite
salt concentration, but its magnitude grows infinitely like $\left(
C_{\infty }^{s+}\right) ^{-0.355}$ (dashed line) as $C_{\infty
}^{s+}\rightarrow 0$. The reason is that the net ion concentration $C\left(
\mathbf{r}\right) =C^{r+}+C^{s+}$ is small near the rear sphere pole in this
case since both $C^{r+}$ and $C^{s+}\ $tend to zero, while the gradient $%
\mathbf{\nabla }C$ is finite. As a result the integrand in (\ref{v_o}) tends
to infinity.

\begin{figure}[ht]
\centering
\includegraphics[width=1\columnwidth]{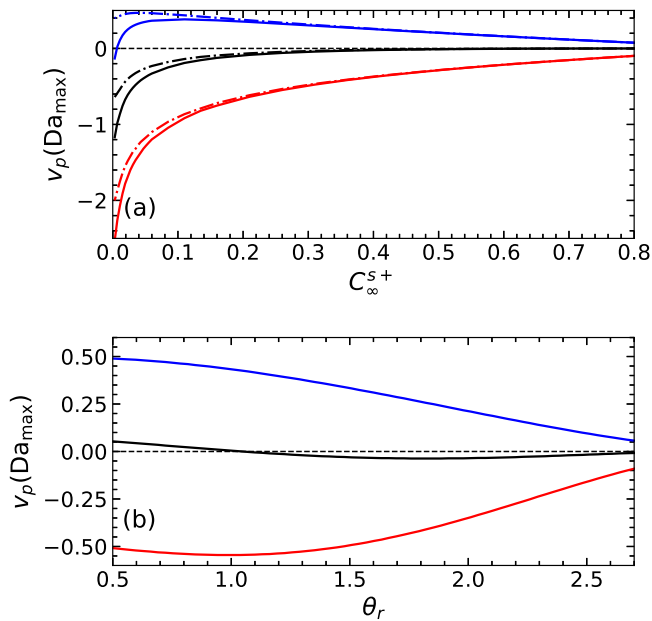}
\caption{Limiting velocities (solid lines) as functions of (a) $C_{\infty }^{r+}$ for $%
\protect\theta _{r}=\protect\pi /2$ and (b) of $\protect\theta _{r}$ for $%
C_{\infty }^{s+}=0.3$. Dashed-dotted lines correspond to the contributions
of the inner region $v_{pi}$.}
\label{vpmax}
\end{figure}

We show the contribution of the inner region $v_{pi}$ evaluated by using
Eqs. (\ref{v_pedl}), (\ref{slip}) and the total particle velocity $%
v_{p}=v_{po}+v_{pi}$ in Fig. \ref{vpmax}. The velocity $v_{pi}$ is dominant
while $v_{po}$ is negligible at finite salt concentration $C_{\infty }^{s+}$
and any $\phi _{s}.$ Up to $C_{\infty }^{s+}\simeq 0.4,$ the velocity $v_{pi}
$ is nearly linear in $C_{\infty }^{r+}$ (see Fig.~\ref{vpmax}(a)), but at
small $C_{\infty }^{s+}$ it grows infinitely, similar to $v_{po}$. The
origin of the singularities for the two contributions is the same: $%
C\rightarrow 0$ near the rear pole in the limit $C_{\infty }^{s+}\rightarrow
0,$ $\mathrm{Da}\rightarrow \mathrm{Da}_{\mathrm{max}}.$ However, the
singularity for $v_{pi}$ is weaker than that for $v_{po}.$ The reason is
that the singularity of the slip velocity (Eq. (\ref{slip})) in the
integrand of (\ref{v_pedl}) is weaker (proportional to $C^{-1})$ than that
of the integrand in Eq. (\ref{v_o}) for $v_{po}$ (proportional to $C^{-3}).$
The contribution of the outer region becomes dominant at very small $%
C_{\infty }^{s+}$, and the total velocity is negative for any $\phi _{s}.$
The maximum particle velocity decays with the angle $\theta _{r}$ i.e. with
the area of the catalytic zone (see Fig.~\ref{vpmax}(b)).

\section{Conclusion}

\label{s5}

We have developed a theory of a self-propulsion of catalytic ionic swimmers
in a media containing both active (released from the swimmer surface) and
passive salt ions. The theory have involved the two types of swimmers, which
release only one type of ions (Type I) or equal fluxes of cations and anions
(Type II). In the limit of a thin EDL, we have derived the analytic solution
for the concentrations of all ionic species at distances comparable to the
particle size (outer region). For the swimmer of Type I, the constraint on
the maximum Damk\"{o}hler number $\mathrm{Da}_{\mathrm{max}}$ follows from the
depletion of active ions near the adsorbing particle side (their
concentration should be positive). $\mathrm{Da}_{\mathrm{max}}$
decreases with the salt concentration $C_{\infty }^{s+}$ but increases with
the area of adsorbing part. The particle propulsion velocity grows
infinitely when $\mathrm{Da}\rightarrow \mathrm{Da}_{\mathrm{max}},\
C_{\infty }^{s+}\rightarrow 0.$

\begin{acknowledgments}

This work was supported by the Ministry of Science and Higher Education of the Russian Federation.
\end{acknowledgments}

\section*{DATA AVAILABILITY}

The data that support the findings of this study are available within the
article.

\bibliographystyle{unsrt}
\bibliography{dph,current}

\end{document}